**High Numerical Aperture Achromatic Meta-Devices through Dispersion Compensation**

*Yuzhong Wang, Axiang Yu, Yayun Cheng, Yongkang Dong, Jiaran Qi\*, Andrea Alù\**

Y. Wang, A. Yu, Y. Cheng, J. Qi

Department of Microwave Engineering, School of Electronics and Information Engineering, Harbin Institute of Technology, Harbin 150001, China.
E-mail: qi.jiaran@hit.edu.cn.

Y. Dong

National Key Laboratory of Laser Spatial Information, Harbin Institute of Technology, Harbin, 150001, China

A. Alù

Photonics Initiative, Advanced Science Research Center, The City University of New York, New York 10031, USA.
E-mail: aalu@gc.cuny.edu.

**Abstract**

Dispersion engineering is a long-standing challenge in optical systems, and it is particularly important for metasurfaces, which naturally suffer from strong chromatic aberrations due to their ultralow profile. Stacks of metasurfaces have recently implemented dispersion control to address these challenges. However, these approaches still suffer from bottlenecks in terms of the available material refractive index and required aspect ratios, resulting in limited phase and group delay coverage, constraining their numerical aperture (NA), size and operating bandwidth. To address these challenges, we explore a dispersion compensation strategy combined with full-wave simulation-free inverse design to implement ultra-high NA, broadband dispersion control in metasurfaces, not requiring large refractive index materials and high aspect ratio processing technology. We experimentally demonstrate multiple meta-devices with highly customized dispersion engineering in the microwave regime, including broadband achromatic diffraction-limited meta-devices with NA=0.98 and 60% fractional bandwidth. Our proposed platform explores a paradigm for dispersion control with metasurfaces, which may facilitate advanced and scalable dispersion functionalities.

# 1. Introduction

Frequency and spatial dispersion arise naturally in any material. Typically, natural materials exhibit positive dispersion features, i.e., in the spatial domain they deflect a larger angle for electromagnetic waves at a higher frequency. For example, the dispersion of a refractive prism, which separates white light into a continuous chromatic spectrum, can be effectively used to implement spectrometers and spectroradiometers. However, in optical sensing systems, these dispersion features may cause challenges in imaging quality and accurate perception.[1] By contrast, diffractive optical devices have material-independent negative dispersion features related to their functionality, which can be expressed as a frequency-dependent function.[2-4] Optical elements with different dispersion properties may cause desirable or undesirable effects in different application scenarios. In general, achromatic optical devices with broadband imaging features are necessary for near-eye displays,[5] cameras,[6] and microscopy,[7] in order to remove image deterioration due to chromatic aberrations. Meanwhile, strong tailored dispersion can be useful to realize high-resolution spectrometers[8-10] and high-performance frequency division multiplexing devices.[11-13] Dispersion-customized control is beneficial for multicolor holography,[14,15] multifunctional frequency-dependent manipulation,[16,17] pulse shaping,[18] and multi-spectral imaging.[19] Therefore, enabling dispersion-customized control adapted to specific systems with various frequency-dependent functionalities is a significant ongoing research field for optics and photonics at large.

Metasurfaces, as ultrathin optical elements consisting of subwavelength meta-atoms with tailored electromagnetic properties, have witnessed vigorous development in recent years. The advent of metasurfaces has prompted the evolution of optical components towards planarization, lightweight and integration. Benefiting from their exotic properties able to efficiently transform the properties of the incoming optical wavefront, including amplitude, phase, polarization, and frequency, a wide variety of metasurfaces with outstanding functionalities have been proposed, such as metalenses,[20-22] all-optical processors,[23-25] holograms,[26-28] beamformers,[29] intelligent adaptive devices,[30] and more. Most of these metasurfaces are operated at one or a few frequencies, and they typically experience unwanted dispersion issues, leading to chromatic aberrations.[31] Considering practical application requirements, achromatic and dispersion-customized control over metasurfaces is an urgent requirement for various imaging systems.[32,33]

Previous works have made efforts to control metasurface dispersion, mainly focusing on multi-dispersion meta-atoms structural design and approaches to synthesize metasurfaces with dispersion-differentiated meta-atoms.[34-53] The desired dispersion control may be achieved with phase and group delay global matching algorithms, searching multi-structure meta-atoms to fit the required dispersion curves.[54] Nevertheless, this approach often leads to challenging requirements on aspect ratio and refractive index in the meta-elements. Considering realistic fabrication and material availability, this limitation results in dispersion-engineered metasurfaces with restricted numerical aperture (NA), size, operating bandwidth and scalability. The folding group delay and phase delay extension approach can alleviate requirements for the aspect ratio and refractive index of meta-atoms, while retaining the desired group delay distribution.[55,56] The dispersion control meta-devices implemented by these methods, however, may have frequency-filtering characteristics, i.e., operating at multiple predefined frequencies, rather than over a broad continuous bandwidth. Approaches such as stacked meta-atoms[57-59] and cascaded metasurfaces[54,60-62] may alleviate the dependence of frequency-related meta-devices on the dispersion modulation capability of meta-atoms with high refractive index and aspect ratio. But again, the stacked meta-atoms approach can only enable achromatic meta-devices at a few discrete frequency points. Previous dispersion-engineered meta-devices designed by cascaded metasurfaces have adopted the ray-tracing technique based on the equal-optical-path principle,[54,60] which may be challenging to satisfy for custom dispersion engineering. Cascaded meta-devices designed by topology optimization methods with the help of finite-difference-time-domain simulations require significant computational resources and sacrifice spatial resolution.[61,62] Nowadays, there are still major challenges in realizing high-NA broadband meta-devices with superior dispersion engineering performance, which are particularly essential in fields that require high resolution and large deflection angles for energy harvesting or emission, such as lithography, spectroscopy, and microscopy.[63-65] Therefore, a general theoretical framework and design methodology, capable of circumventing restrictions on processing technology and material properties, to efficiently synthesize high-NA broadband meta-devices with highly customized dispersion control, is currently not available.

Here, we introduce a dispersion compensation strategy free from reliance on complex dispersion response meta-atoms and processing technologies, aimed at designing high-NA broadband meta-devices. We present the theoretical framework based on spatially cascaded broadband phase-only modulation components with a constant group delay, which provides a rational perspective to design multi-layer spatially cascaded dispersion control in optical meta-devices. Exploiting this layer-to-layer frequency-dependent response, the phase and group delay of meta-devices are demonstrated with the assistance of a full-wave simulation-free inverse design method, breaking the dispersion constraint of individual meta-atoms. This approach provides a general and efficient pathway to design highly customized dispersion engineered meta-devices without the limitation on NA, operating bandwidth and size. As a proof of concept, we numerically and experimentally demonstrate multiple meta-devices with dispersion control functionalities in the microwave band (7GHz-13GHz, NA=0.87), including achromatic, hyper dispersion, positive dispersion, and segmented dispersion. Meanwhile, a broadband achromatic meta-device with an ultra-high NA of 0.98 and dispersion-tailored beam deflection is experimentally demonstrated, demonstrating the flexibility and validity of the proposed technique. This dispersion compensation strategy offers a universal platform to design dispersion-customized modulation meta-devices.

## 2. Results

Dispersion engineering aims at realizing optical elements that cater to desired phase profiles at several frequency points or over broad bandwidths. As a classical example, consider the design of a diffractive metalens with a frequency-dependent phase profile:

$$\varphi(r, f, d(f)) = -\frac{2\pi f}{c}\left(\sqrt{r^2 + d(f)^2} - d(f)^2\right) - C(f) \tag{1}$$

where $r$, $f$, $c$, $d(f)$, and $C(f)$ are the radial coordinate, frequency, light speed, frequency-dependent focal length, and frequency-dependent constant reference phase, respectively. The phase profile $\varphi(r, f, d(f))$ shows that, for a given spatial coordinate $r$, dispersion engineering is required to support specific group delays, i.e., $\partial \varphi(r, f, d(f))/\partial f$, to enable a desirable phase response at various frequencies. Specifically, in order to obtain an achromatic metalens, with $d(f)$ independent of frequency, the desired phase profile should vary with frequency and its corresponding group delay should change with the spatial coordinate, as shown in **Figure S1**. As the size, operating bandwidth and NA of the designed achromatic metalens increase, more stringent phase and group delay requirements emerge. Rational expressions of the required conditions are shown in **Supplementary Information S1**.

It is however challenging to implement broadband dispersion control for a metalens with a large NA and large size, as it requires meta-atoms with a large range of group delays and phase modulation features. Specifically, as shown in **Figure S1d**, when designing a larger size and larger NA broadband dispersion-related meta-device, the meta-atom library requires a larger range of group delays with identical phase responses at each frequency of interest. Typically, for an achromatic metalens it is possible to obtain meta-atoms with larger group delays adopting higher refractive index materials and more sophisticated structures. However, due to the limited range of phase dispersion that can be achieved by meta-atoms, a single-layer metalens imposes a severe trade-off between size, operating bandwidth and NA.[42] Due to further constraints imposed by material properties and processing techniques, unconventional dispersion responses are challenging, especially for large NA, size, broad bandwidths, and customized dispersion control.

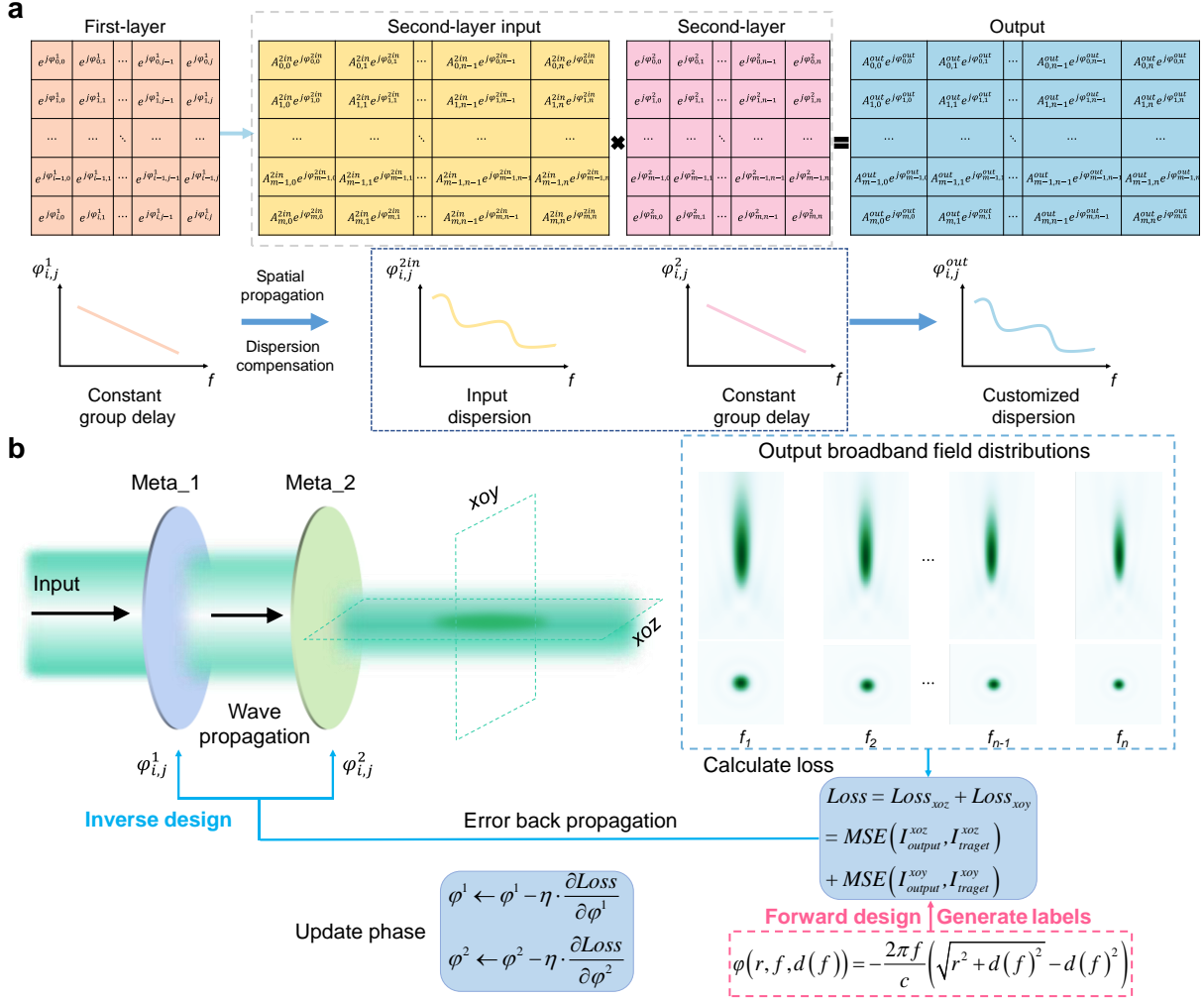

**Figure 1.** Schematic of broadband dispersion engineering in a metasurface. a) Theoretical framework and dispersion compensation architecture. b) Inverse design framework for dispersion-engineering meta-device with focusing functionality.

To enable customized dispersion engineering in metasurfaces overcoming the trade-off between NA, size and operating bandwidth, we explore a dispersion compensation approach utilizing bilayer spatially cascaded architecture. A bilayer platform is introduced to overcome the bottleneck in the single-layer dispersion control scheme, in which dispersion-rich meta-atoms, attained through large refractive index materials and high aspect ratio processing technology, are usually exploited to cater to the demand for stringent group delay and phase coverage. This bilayer architecture presents a novel solution for dispersion engineering, converting the classical approach of designing dispersion control meta-atoms into the procedure of optimizing the phase distribution of bilayer linear-dispersion phase-only metasurfaces.

**Figure 1a** illustrates the underlying theoretical framework of the proposed dispersion compensation approach based on a meta-device consisting of spatially cascaded bilayers with phase-only modulation components, $\varphi_{i,j}^1$ and $\varphi_{m,n}^2$. Each component is a frequency-dependent phase-only modulation factor with constant group delay. Dispersion-customized control introduced by this dispersion compensation architecture aims at realizing tailored phase and group delay features through linear-dispersion phase-only modulation factors. Firstly, the input broadband plane electromagnetic waves modulated by the first-layer frequency-dependent complex-value transmission factors $e^{\varphi_{i,j}^1(f)}$ with linear-dispersion phase-only properties propagate forward in free space to the second layer. The dispersion property of second-layer frequency-dependent complex-value input electric fields $A^{2in}(f)\varphi_{m,n}^{2in}(f)$ breaks away from the original linear-dispersion features by exploiting the additive effect of the complex-valued light fields in spatial propagation. This opens up possibilities for customized dispersion engineering, which is manifested by the fact that the input light fields of the second layer do not have the original linear-dispersion characteristics of the first layer. Next, through rationally manipulating the first-layer linear-dispersion phase-only modulation factors, $\varphi_{m,n}^{2in}$ can be skillfully engineered, which is critical to attaining dispersion-tailored. Utilizing this adjustable input dispersion features $\varphi_{m,n}^{2in}(f)$ modulated by the first layer, by introducing the second-layer linear-dispersion modulation factors $\varphi_{m,n}^2$, the frequency-dependent output phase response $\varphi_{m,n}^{out}(f)$ can be expressed as:

$$\varphi_{m,n}^{out}(f) = \arctan\left( \frac{\sum_{i,j \in A}\left(\tan\left(2\pi f r_{i,j} + \varphi_{i,j}^1 + \varphi_{m,n}^2\right) - 2\pi f r_{i,j}\right)}{\sum_{i,j \in A}\left(1 + \tan\left(2\pi f r_{i,j} + \varphi_{i,j}^1 + \varphi_{m,n}^2\right) \cdot 2\pi f r_{i,j}\right)} \right) \tag{2}$$

where $A$ denotes the total discretized grids in the first-layer, $r_{i,j}$ is the spatial distance between the phase-only modulation factor $\varphi_{i,j}^1$ in the first-layer and the phase-only modulation factor $\varphi_{m,n}^2$ in the second-layer, and $f$ is the working frequency. Desired group delay profiles, deciding on the output dispersion response, are mainly determined by the additive effects of the light field spatial propagation from the first-layer linear-dispersion phase-only modulation factors $\varphi_{i,j}^1$. Output phase profiles, resulting in functionalities of focusing or holography, are mainly dependent on phase compensation of the second-layer linear-dispersion phase-only modulation factors $\varphi_{m,n}^2$. Details of the derivation process regarding equation (2) and some dispersion-tailored examples are provided in **Supplementary Information S2**. Through this dispersion compensation framework, the second-layer complex-value input electric fields $A^{2in}(f)\varphi_{m,n}^{2in}(f)$ at each position ($m$, $n$) is determined by all modulation factors $\varphi_{i,j}^1$ of the first layer, and it converts the original linear-dispersion response into a non-linear one, as shown in **Figures S4a** and **b**. This modification of the original linear-dispersion of the first layer opens the potential for customized output phase and group delay response. To further demonstrate that the frequency-dependent trajectory of **Eq. (2)** can be engineered at will, we exploit the gradient descent algorithm to tailor the output dispersion of a particular phase modulation factor $\varphi_{0,0}^2$ in the second layer. At a fixed spatial distance $r_{i,j}$, by optimizing the linear-dispersion phase-only modulation factors $\varphi_{i,j}^1$ and $\varphi_{0,0}^2$, we can demonstrate large dispersion control, cumbersome to implement by designing meta-atom structures, as shown in **Figures S4c** and **d**. As a result, the output dispersion can be customized to fit desired linear or nonlinear phase profiles and group delays across broad bandwidths through dispersion compensation strategy.

To enable dispersion-engineered meta-devices through the dispersion compensation approach, full-wave simulation-free inverse design is adopted to correlate the desired frequency-dependent output fields with an optimized bilayer. This inverse design framework aims to straightforwardly obtain bilayer phase profiles oriented by the functionality of the desired dispersion-engineered meta-devices, realizing the end-to-end design flow, without the tricky group delay matching. The complete design process is illustrated in **Figure 1b**. Firstly, for dispersion-engineered meta-devices with the desired focusing functionality, the forward design is exploited to generate frequency-dependent target field distributions in the *xoy* and *xoz* planes as labels. Next, randomly initialized phase modulation factors are selected in the bilayer. Frequency-dependent output electric field distributions $\vec{E}(r_{m,n}, f)$ on the target position under plane wave incidence can be expressed as:

$$\vec{E}(r_{m,n}, f) = \sum_{m,n \in B} \left( H(r_{m,n}, f) \exp(j\varphi_{m,n}^2) \left( \sum_{i,j \in A} H(r_{i,j}, f) \exp(j\varphi_{i,j}^1) \right) \right) \qquad (3)$$

where $H(r_{m,n}, f)$ is the Rayleigh–Sommerfeld diffraction transfer function shown in **Eq. (S4)**, $r_{m,n}$ is the spatial distance between the target position and the ($m$, $n$)-th phase modulation factor in the second layer, $B$ denotes the total discretized grids in the second layer. The Adam optimizer is used to update trainable parameters and the mean-squared-error (MSE) loss function is used to evaluate errors between the output and target fields. The gradient descent algorithm and error back-propagation algorithm are used to minimize losses until the model converges. The desired dispersion-engineered meta-devices consisting of bilayer linear-dispersion phase-only modulation factors can be obtained efficiently through this inverse design process.

To verify the feasibility of our approach, we first demonstrate an ultra-high NA achromatic metasurface (NA=0.98), with aperture radius of 150mm and focal length of 30mm, at the working frequency range spanning from 7GHz to 13GHz. Broadband meta-atoms, as phase-only modulation factors, with a constant group delay are used. Detailed structures and dispersion responses of the meta-atom with a period of 6mm are given in **Supplementary Information S3**. The meta-device consists of two phase-only metasurfaces with a distance of 100mm. Simulated results of seven frequencies in the working range, shown in **Figure S6,** display excellent achromatic capability of the designed meta-device with a high NA. In the optimization process, we have chosen only a few integer frequency points with intervals of 1GHz as inputs to alleviate computational resource occupation and time-consuming optimization. In this case, it takes only a few minutes to perform 3000 iterations on a personal computer. For further validation, **Figure S7** shows the simulation of the meta-device at other non-preoptimized frequencies, and similarly exhibits outstanding achromatic effects. These results show that the obtained meta-device can control the dispersion response within a broadband range, rather than a few specific frequency points. By PCB processing technology, as detailed in the "Methods" section, a sample prototype of the ultra-high NA achromatic meta-device was fabricated. The experimental characterization results are shown in **Figure 2**. The detailed measurement process is described in the "Methods" section, the schematic diagram of the experimental setup is shown in **Figure 2a**, and practical images are shown in **Supplementary Information S9**. The experimental axial plane and focal plane intensity distributions at seven different frequencies are shown in **Figure 2b**. The experimental and simulated focal length, with negligible deviation not exceeding 5% of the central wavelength, of the proposed ultra-high NA achromatic meta-device, compared with the focal length of the normal chromatic metalens at seven different frequencies is shown in **Figure 2c**. The focusing efficiency (FE) is defined as the power collected at the focal spot with a radius 1.5 times the full width at half maximum (FWHM) divided by the sum of the incident power. The measured FE of the achromatic meta-device compared with the simulated FE is shown in **Figure 2d**. The measured FE for each frequency is around 25%, which is slightly lower than the simulated FE, and gradually increases with frequency. It should be noted that the reason for the low FE of the designed meta-device is not due to the design method but to the NA of the metasurface. As shown in **Figure S6d**, in contrast to the simulated FE of seven traditional diffraction-limited metalenses with same size and focal length designed respectively at each frequency, the proposed achromatic meta-device has a close or even higher simulated FE. The measured and simulated FWHM of the designed ultra-high NA achromatic meta-device is shown in **Figure 2e**, compared with the theoretical diffraction limited FWHM (0.51λ/NA). This result shows that the resolution of the proposed ultra-high NA achromatic meta-device is close to the theoretical diffraction limit within the working frequency range and with acceptable levels of FE. Experimental results agree with simulated results and slight deviations between them may be attributed to processing or measurement misalignment. To analyze these factors, simulated experiments were carried out to observe the impact of misalignment and manufacturing errors on the performance of the fabricated achromatic meta-device with NA of 0.98. Among these misalignment errors, the tilt and lateral displacement have a significant influence on the focusing position and FE of the focal spot. Rotational errors have slight effects on the normalized field distribution and FE due to rotational symmetry of the focusing device, while longitudinal displacements have visible effects on them. On the other hand, the performance is acceptably influenced by random manufacturing errors, with only a slight attenuation on FE, and the processing technology employed is mature enough that large areas of processing errors are essentially non-existent. Although misalignment errors are the main cause of device performance degradation in cascaded systems, the proposed meta-device shows excellent error tolerance within correctable displacement ranges. More detailed discussions are shown in **Supplementary Information S8**.

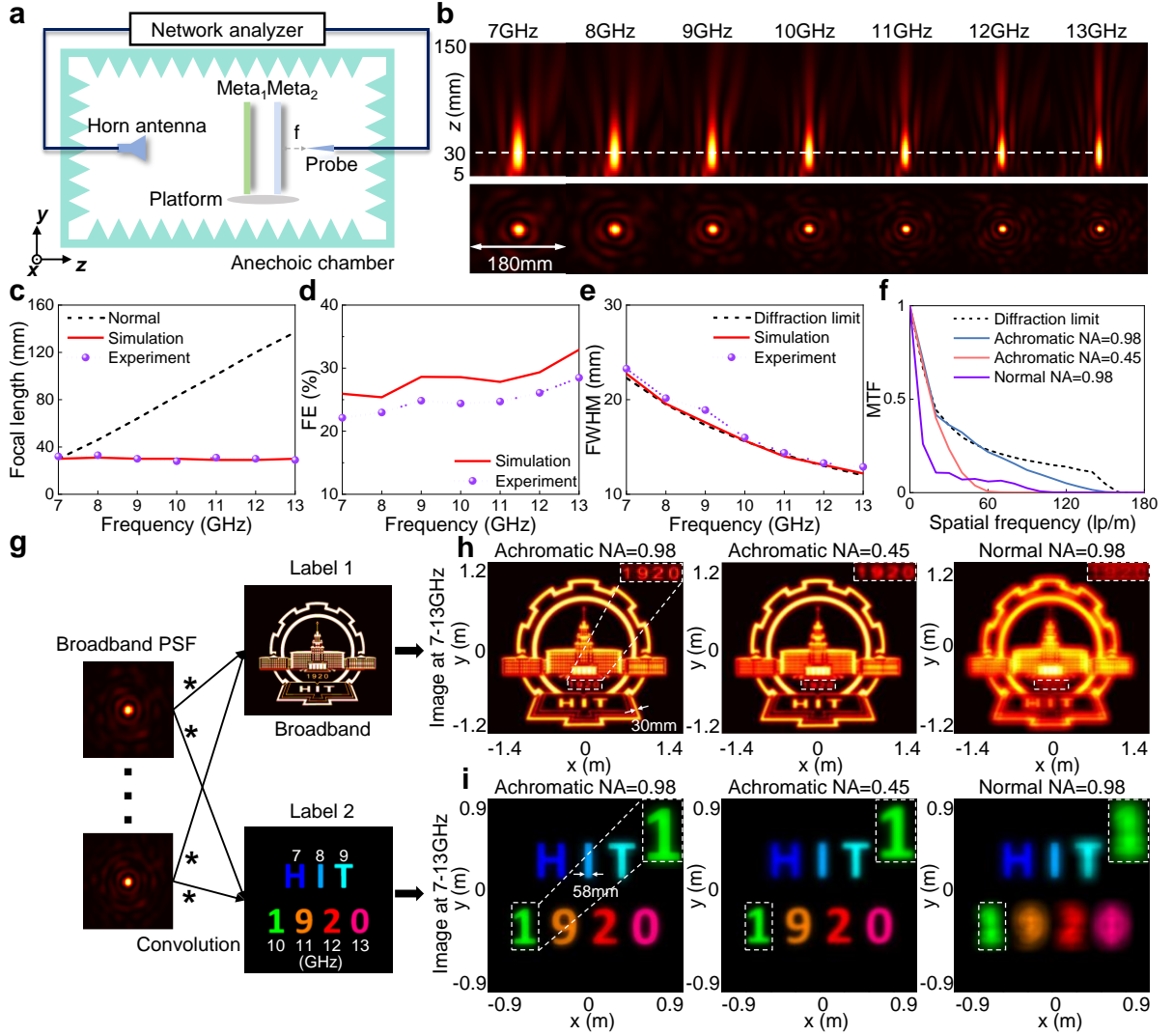

**Figure 2.** Performance characterization of the proposed broadband achromatic meta-device with NA=0.98. a) Schematic of the measurement configuration. b) Experimental normalized intensity distributions of the axial plane and focal plane. c) Experimental and simulated focal length of the proposed broadband achromatic meta-device compared with the focal length of the normal negative chromatic metalens. d) Measured and simulated FE. e) Measured and simulated FWHM compared with the theoretical diffraction limitation FWHM. f) Broadband MTF compared with the diffraction limitation MTF. g) Broadband computational imaging process. h) Broadband computational imaging results of label 1 under a broadband waves incidence. i) Colored computational imaging results of label 2 under a broadband waves incidence.

For demonstration purposes, the size of the realized meta-device is only ten times the central working wavelength, which is mainly due to the fact that available manufacturing techniques cannot process a metasurface of large size in the microwave band. Moreover, the operating bandwidth of meta-atoms chosen for validation is finite, with the transmission amplitude dropping rapidly outside 7-13 GHz. It should be noted that this method can be applied to enable dispersion engineering of meta-devices with a wider bandwidth and a larger size. This is due to the fact that, as the size increases, the degrees of freedom available for dispersion compensation in the first phase layer also increase, which in turn provides greater dispersion modulation capabilities to the frequency-dependent output phase response of the second phase layer.

To demonstrate this opportunity, a larger-size high-NA broadband achromatic meta-device (NA=0.99), whose size is 50 times the central working wavelength and the focal length is 100mm, was numerically designed using the same approach. Numerical results are shown in **Figure S8**. The achromatic effect of the large-size meta-device shows the same outstanding performance. Furthermore, we have also numerically demonstrated another broadband achromatic meta-device with an NA of 0.98 operating at 7-21 GHz. Numerical results are shown in **Figure S9**. It similarly shows excellent achromatic properties over a wider operating bandwidth. However, compared to the achromatic meta-device at 7-13 GHz, in the low-frequency region, it has a lower FE and a somewhat sacrificed but acceptable resolution relative to the diffraction limitation. Essentially, this is a performance trade-off between operating bandwidth, FE and resolution in the optimization process. The performance of the obtained meta-devices may not be necessarily optimal, but their resolution and efficiency are close to the diffraction limit, enabling them to be applied to high-quality achromatic imaging. Moreover, at a fixed size and NA, spatial resolution and efficiency may be further tailored by adjusting preset labels on demand. We also designed achromatic meta-devices with NA of 0.87, 0.71, 0.6, and 0.45. Simulated results are shown in **Figure S10-S13**, which all exhibit excellent achromatic capabilities and a resolution close to the diffraction limit. These results demonstrate that the desired group delays and phase response can be implemented, enabling achromatic meta-devices with strong performance.

To further evaluate the imaging performance, we acquired the broadband modulation transfer function (MTF) of the achromatic meta-device with NA=0.98, the achromatic meta-device with NA=0.45, and the normal negative chromatic metalens with NA=0.98 designed at 7GHz, and compared with the theoretical diffraction limited MTF with NA=0.98 at 13GHz. As shown in **Figure 2f**, the proposed high-NA achromatic meta-device exhibits larger resolution and contrast, close to the theoretical diffraction limit for NA=0.98 at 13GHz. By contrast, the conventional metalens with NA=0.98 designed at 7GHz, while showing a good focus has a worse broadband contrast in the low spatial frequency region, due to negative chromatic aberration. As shown in **Figure 2g**, we numerically calculated broadband imaging results by convolving the point spread function (PSF) with the label image at different frequencies. Label 1 with the size of 2.8m×2.4m is illuminated by broadband electromagnetic waves from 7-13 GHz, and broadband imaging for the three meta-devices are shown in **Figure 2h**. The high-NA achromatic meta-device performs the best imaging result, clearly distinguishing a linewidth of 30mm, close to the theoretical diffraction limit of 28.7mm for NA=0.98 at 13GHz (1.22$\lambda$/NA). The image result of the lower NA achromatic meta-device is blurred due to its lower resolution, yet it still shows a distinct profile for the large linewidth, benefiting from its good broadband imaging capability. The image result of the negative chromatic metalens displays obvious frequency blurring effects, which generate stray images around the outline. Label 2 with the size of 1.8m×1.8m consists of seven letters illuminated by electromagnetic waves at different frequencies. For ease of observation, we have colored each letter. The high-NA achromatic meta-device can clearly image letters for each frequency. The low-NA achromatic meta-device shows good broadband imaging with some resolution-related edge blurs due to the lower NA. However, the image of the negative chromatic metalens shows significant frequency-related blurring at non-designed frequency points.

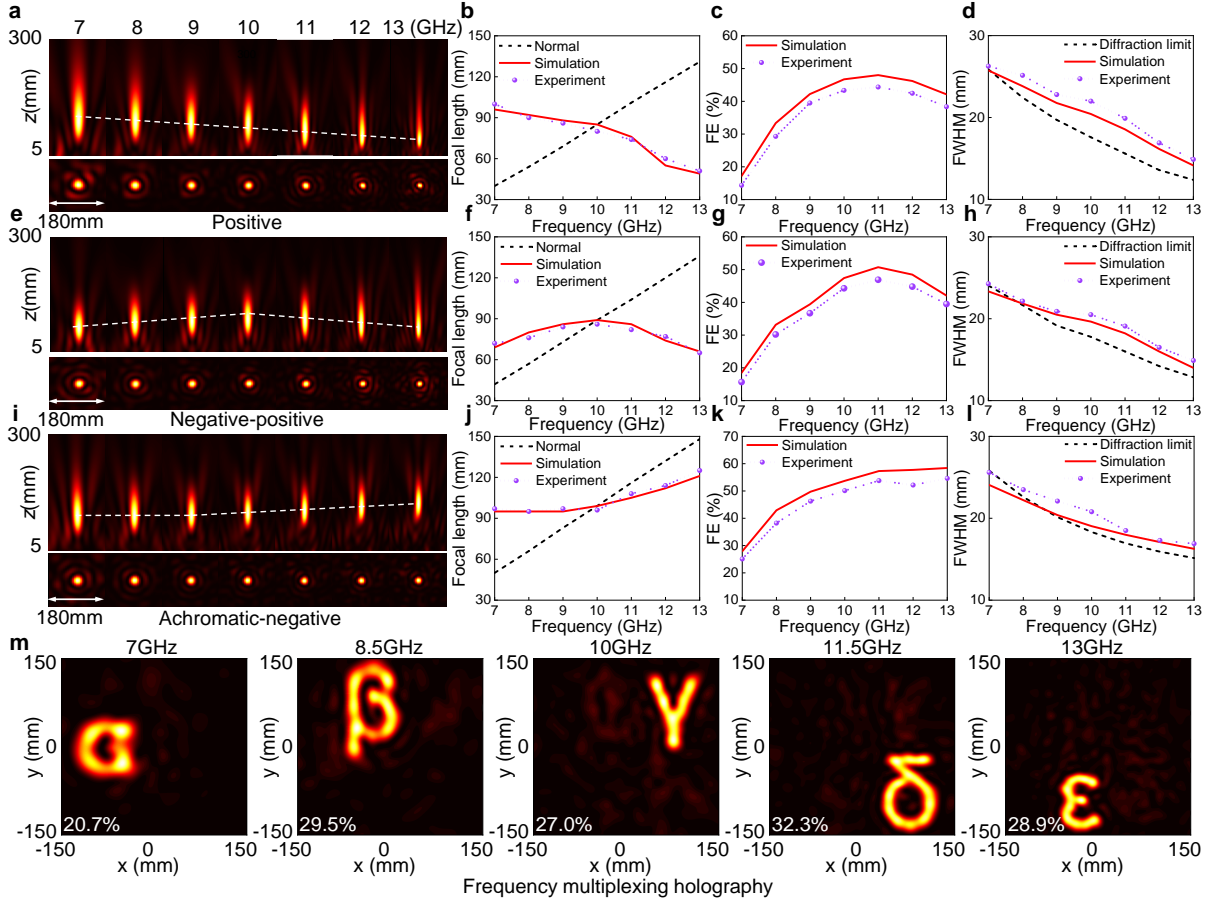

**Figure 3.** Performance characterization of dispersion-engineered meta-devices. a) Experimental normalized intensity distributions of the axial plane and focal plane, b) focal length, c) FE, and d) FWHM of the positive dispersion meta-device. e) Experimental normalized intensity distributions of the axial plane and focal plane, f) focal length, g) FE, and h) FWHM of the segmented negative-positive dispersion meta-device. i) Experimental normalized intensity distributions of the axial plane and focal plane, j) focal length, k) FE, and l) FWHM of the segmented achromatic-negative dispersion. m) Simulated results of the frequency multiplexing holography meta-device.

To further characterize the proposed dispersion engineering approach, we designed and fabricated three high-NA meta-devices with highly customized dispersion features, i.e., positive dispersion, segmented negative-positive dispersion, and segmented achromatic-negative dispersion, respectively. Three meta-devices, all consisting of two phase-only metasurfaces with a distance of 100mm, have an aperture radius of 150mm and operate from 7GHz to 13GHz. Experimental results of the three fabricated meta-devices are shown in **Figure 3**. As shown in **Figure 3a**, the first meta-device has a positive dispersion, where the focal length varies from 95mm (NA=0.84) at 7GHz to 49mm (NA=0.95) at 13GHz. As shown in **Figure 3e**, the second meta-device has a segmented dispersion, where a weak negative dispersion has a focal length from 70mm (NA=0.91) at 7GHz to 90mm (NA=0.86) at 10GHz and a positive dispersion has a focal length from 90mm at 10GHz to 65mm (NA=0.92) at 13GHz. As shown in **Figure 3i**, the third meta-device also has a segmented dispersion effect, which exhibits achromatic with a 95mm focal length from 7GHz to 9GHz and a weak negative dispersion with a focal length from 95mm at 9GHz to 120mm (NA=0.78) at 13GHz. The experimental intensity distributions of the three prototypes all have good agreement with the simulated ones shown in **Figure S14-S16**. The dependence of experimental and simulated focal lengths with frequency of the three fabricated meta-devices, compared with the ones of conventional chromatic metalens, are shown in **Figure 3b, f,** and **j**. These results display excellent dispersion control of the proposed meta-device, meeting the desired dispersion trends. The comparison of the measured and simulated FE of three meta-devices is shown in **Figure 3c, g, and k**. The measured FE is slightly decreased within acceptable levels, but has an identical variation trend with the simulated one, which may be due to manufacturing and measurement errors. Meanwhile, the measured and simulated FWHM of the focus spots generated by the three meta-devices at specific frequencies and focal planes compared with the diffraction limitation FWHM is shown in **Figure 3d, h,** and **l**. There is a slight but acceptable decrease in resolution for the custom dispersion meta-devices. In addition to the above three meta-devices, we have also designed several meta-devices with other dispersion control capabilities, whose simulated results are as shown in **Supplementary Information S6**, including hyper negative dispersion, weak negative dispersion, and other segmented dispersion features. We have also designed a frequency multiplexing holography meta-device, which displays α at 7GHz, β at 8.5GHz, γ at 10GHz, δ at 11.5GHz, and ε at 13GHz. Simulated results are shown in **Figure 3m** and the hologram efficiency is reported in the graph. All results indicate that the dispersion compensation strategy jointly with the inverse design method allows dispersion-engineered meta-devices to be successfully engineered with optimal performance.

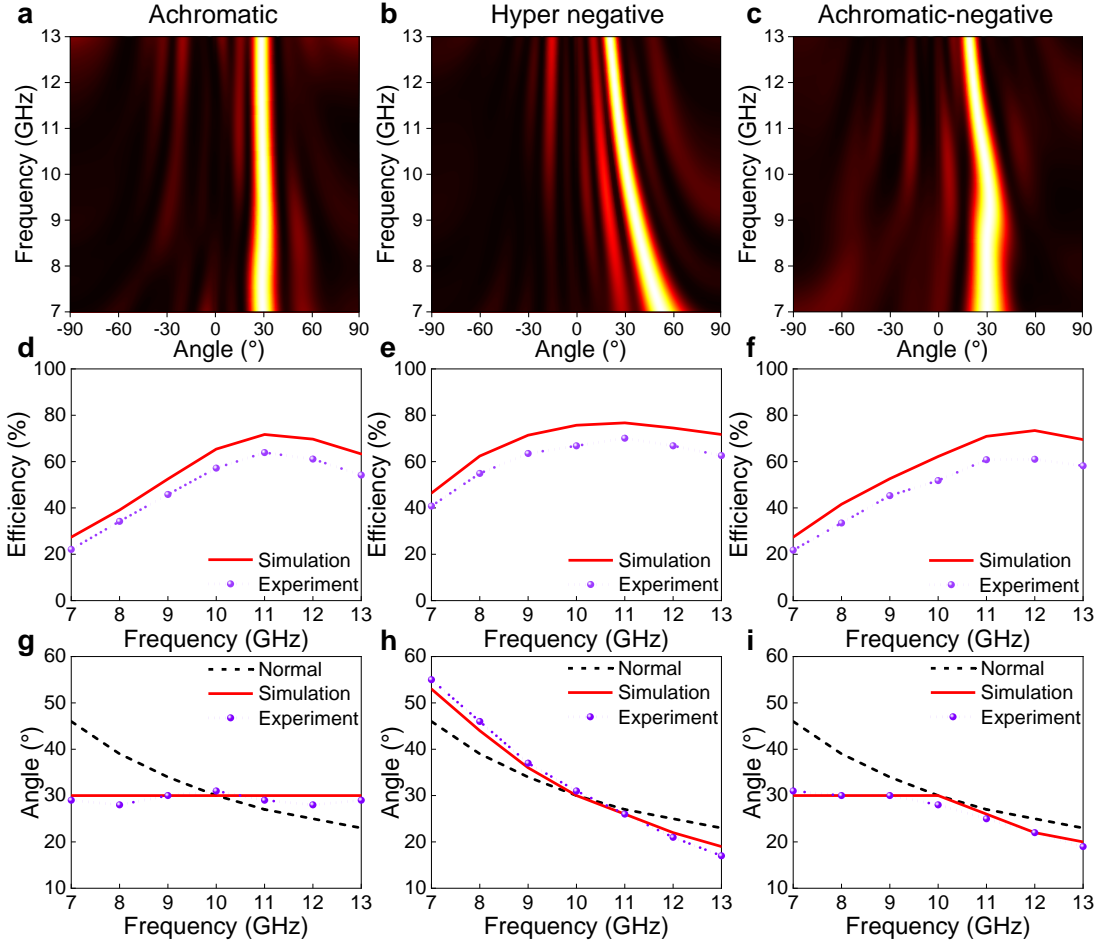

**Figure 4.** Performance characterization of the dispersion-engineered beam deflection meta-devices. a) Experimental far-field normalized intensity distribution, d) beam deflection angle, and g) efficiency of the 30° achromatic beam deflection meta-devices. b) Experimental far-field normalized intensity distribution, e) beam deflection angle, and h) efficiency of the hyper-negative beam deflection meta-devices. c) Experimental far-field normalized intensity distributions, f) beam deflection angle, and i) efficiency of the segmented achromatic-negative beam deflection meta-devices.

The proposed approach may also be used to enable broadband far-field beam deflection meta-devices, which have been widely deployed in the antenna and communication fields, in order to tailor frequency sweep according to practical application requirements. In order to enable dispersion-engineered beam deflectors, the frequency-dependent output far-field function $F(\theta, \varphi, f)$ in the optimization process can be expressed as:

$$F(\theta,\varphi,f) = \sum_{m,n \in B} \left( U(m,n,\theta,\varphi,f) \exp(j\varphi_{m,n}^2) \left( \sum_{i,j \in A} H(r_{i,j},f) \exp(j\varphi_{i,j}^1) \right) \right) \qquad (4)$$

where $U(\theta, \varphi, f)$ is the far-field transfer function:

$$U(m,n,\theta,\varphi,f) = \exp\left(j2\pi f\left[m\sin(\theta)\cos(\varphi) + n\sin(\theta)\sin(\varphi)\right]\right) \tag{5}$$

where $m$ and $n$ are the coordinate positions of the phase factor in the second layer, $\theta$ is the elevation angle and $\varphi$ is the azimuth angle. After applying the same optimization process shown in **Figure 1b**, far-field beam deflection meta-devices with highly customized dispersion can be designed. As a proof of concept, we experimentally demonstrated three dispersion-tailored control beam deflectors, including achromatic, segmented achromatic-negative combinations, and hyper-negative dispersion. All dispersion-engineered beam deflectors consist of two-layer spatially cascaded metasurfaces with a distance of 100mm. The size of each metasurface is 180mm×180mm. Experimental far-field normalized intensity distributions for the three mete-devices are shown in **Figure 4a-c**. The direction of maximum beam energy at all frequency points is consistent with preset angles, displaying excellent far-field dispersion control capability, as shown in **Figure 4d-f**. The experimental results are in good agreement with the simulation results shown in **Figure S22**. Here, using antenna theory, we define the beam deflection efficiency as the actual beam intensity at the preset direction divided by the theoretical beam intensity generated by the same size aperture. Experimental and simulated efficiencies of three beam meta-devices at each frequency are shown in **Figure 4g-i**. We also simulated other interesting dispersion-tailored beam deflectors, as shown in **Supplementary Information S7**, including single-beam achromatic, single-beam segmented dispersion control, and multi-beam dispersion-engineered meta-devices. All results confirm that the proposed dispersion engineering enabled by dispersion compensation strategy jointly with inverse design can efficiently enable customized beam deflection control over broad bandwidths.

## 3. Discussion

In this paper we have introduced a dispersion compensation strategy that provides a general architecture to enable broadband dispersion-customized meta-devices without restriction on NA, operating bandwidth and size. The desired phase response and group delays, which are difficult to obtain through meta-atom structure design, can be readily accessed over broad bandwidths by gradient-descent-based optimization of dual-layer spatially cascaded phase-only metasurfaces. This approach can be scaled to higher frequency bands, due to the easy-to-implement phase required for the two layers. With the aid of full-wave simulation-free inverse design, dispersion-engineered meta-devices can be easily realized without the requirement of sophisticated broadband phase and group delay matching algorithms. As a proof-of-concept demonstration, we have realized broadband at-will dispersion control meta-devices, including high-NA metalenses and beam deflectors, in the microwave band. Experimental and simulated results demonstrate the feasibility and reliability of the proposed method to enable dispersion-customized meta-devices. The proposed dispersion compensation strategy may facilitate advanced dispersion-engineered meta-devices and their potential applications in broadband high-NA imaging and spectroscopy applications.

To highlight the advantages of the proposed method, a performance comparison and detailed discussion with other representative broadband dispersion-engineered meta-devices are shown in **Supplementary Information S10**, showcasing significant improvements in NA, achromatic resolution and dispersion control. In terms of the underlying principles, the proposed dispersion compensation framework eliminates the necessity to rely on the meta-atom structural dispersion properties, which is the fundamental difference from conventional dispersion control approaches. It significantly breaks through the performance constraints and challenges of existing dispersion-tailored metasurfaces, resulting from the accessible limited material refractive index and manufacturing aspect ratios. Only requiring to employ regular linear-dispersion phase-only metasurfaces with the assistance of the dispersion compensation mechanism, it is feasible to implement dispersion-engineered meta-devices, without restrictions in terms of NA, size and operating bandwidth. In terms of design methodology, the proposed full-wave simulation-free inverse design can directly correlate the meta-device functionality to the metasurface parameters, enabling an end-to-end efficient design process, without the complicated group delay matching algorithms. This procedure ensures high performance of the achromatic meta-device, i.e., high NA diffraction-limited focusing and dispersion-tailored capabilities.

## 4. Experimental Section

*Sample fabrication:* The prototypes of meta-devices proposed in this work are fabricated by the traditional printed circuit board (PCB) technique. In the manufacturing process, the metasurface is composed of two-layer dielectric structures and three-layer metal. The dielectric substrate with 3 mm thickness has a relative permittivity $\varepsilon_r = 2.65$. The copper layer has a thickness of 0.035 mm. The total thickness of the metasurface sample is 6.105mm, which is approximately 0.2 wavelength at 10GHz.

*Experimental measurements:* Experimental measurements are performed in the microwave chamber to prevent interference from ambient electromagnetic waves. A wideband horn antenna as the feeding source emits linearly polarized plane electromagnetic waves. For near-field measurement experiments, we adopted a rectangular waveguide to receive the amplitude and phase of the electromagnetic wave on the output plane. The waveguide probe is controlled by a scanning device to obtain the electric fields over the entire plane. For far-field measurement experiments, we adopt the same receiving and transmitting broadband horn antennas. The fabricated meta-devices and transmitting antenna are placed on an automated turntable to provide angular scanning, and the receiving antenna is fixed in place, which meets the far-field test conditions. In both measurement experiments, the vector network analyzer is connected to transmitting and receiving antennas to obtain the amplitude and phase of the electromagnetic information to be measured.

**Supporting Information**

Supporting Information is available from the Wiley Online Library or from the author.

**Acknowledgements**

Y. W., A. Y., Y. C., J. Q. were supported by the National Natural Science Foundation of China under Grant 62271170, Grant 62371159, Grant 62071152, and Grant 61901242. Y. D. was supported by the National Key Laboratory of Laser Spatial Information Foundation. A. A. was supported by the Simons Foundation.

**Author contributions:** Y. W., J. Q., A. A. conceived the idea. Y. W., A. Y. performed the sample fabrication and performed the experiment. Y. W., A. Y., Y. C., Y. D., J. Q., A. A. contributed to write the manuscript. J. Q. supervised the study.

**Competing interests:** The authors declare that they have no competing interests.

**Data and materials availability:** The data that support this work are available from the corresponding author upon reasonable request.

<div style="text-align: right;">
Received: ((will be filled in by the editorial staff))
Revised: ((will be filled in by the editorial staff))
Published online: ((will be filled in by the editorial staff))
</div>

**References**

[1]    M. Chen, Y. Wu, L. Feng, Q. Fan, M. Lu, T. Xu, D. P. Tsai, *Adv. Optical Mater.* **2021**, *9*, 2001414.

[2] M. Pan, Y. Fu, M. Zheng, H. Chen, Y. Zang, H. Duan, Q. Li, M. Qiu, Y. Hu, *Light Sci. Appl.* **2023**, *12*, 17.

[3] X. Xiao, Y. Zhao, X. Ye, C. Chen, X. Lu, Y. Rong, J. Deng, G. Li, S. Zhu, T. Li, *Light Sci. Appl.* **2022**, *11*, 323.

[4] W. Chen, A. Zhu, F. Capasso, Flat optics with dispersion-engineered metasurfaces. *Nat. Rev. Mater.* **2020**, 5, *8*.

[5] W. Zang, Q. Yuan, R. Chen, L. Li, T. Li, X. Zou, G. Zheng, Z. Chen, S. Wang, Z. Wang, S. Zhu, *Adv. Mater.* **2020**, *32*, 1904935.

[6] X. Zou, G. Zheng, Q. Yuan, W. Zang, R. Chen, T. Li, L. Li, S. Wang, Z. Wang, S. Zhu, *PhotoniX*, **2020**, *1*, 2.

[7] J. Yao, R. Lin, M. Chen, D. P. Tsai, *Adv. Photon.* **2023**, *5*, 024001.

[8] R. Wang, M. A. Ansari, H. Ahmed, Y. Li, W. Cai, Y. Liu, S. Li, J. Liu, L. Li, X. Chen, *Light Sci. Appl.* **2023**, *12*, 103.

[9] C. Chen, W. Song, J.-W. Chen, J.-H. Wang, Y. Chen, B. Xu, M.-K. Chen, H. Li, B. Fang, J. Chen, H. Kuo, S. Wang, D. Tsai, S. Zhu, T. Li, *Light Sci. Appl.* **2019**, *8*, 99.

[10] G. Cai, Y. Li, Y. Zhang, X. Jiang, Y. Chen, G. Qu, X. Zhang, S. Xiao, J. Han, S. Yu, Y. Kivshar, Q. Song, *Nat. Mater.* **2023**. https://doi.org/10.1038/s41563-023-01710-1.

[11] J. Li, X. Li, N. T. Yardimci, J. Hu, Y. Li, J. Chen, Y.-C. Hung, M. Jarrahi, A. Ozcan, *Nat. Commun.* **2023**, *14*, 6791.

[12] M. Cotrufo, A. Arora, S. Singh, A. Alù, *Nat. Commun.* **2023**, *14*, 7078.

[13] L. Zhang, M. Chen, W. Tang, J. Dai, L. Miao, X. Zhou, S. Jin, Q. Cheng, T. Cui, *Nat. Electron.* **2021**, *4*, 3.

[14] Y. Intaravanne, R. Wang, H. Ahmed, Y. Ming, Y. Zheng, Z. Zhou, Z. Li, S. Chen, S. Zhang, X. Chen, *Light Sci. Appl.* **2022**, *11*, 302.

[15] S. So, J. Kim, T. Badloe, C. Lee, Y. Yang, H. Kang, J. Rho, *Adv. Mater.* **2023**, *35*, 2208520.

[16] X. Li, Q. Chen, X. Zhang, R. Zhao, S. Xiao, Y. Wang1, L. Huang, *Opto-Electron Adv*. **2023**, *6*, 220060.

[17] M. Huang, B. Zheng, R. Li, X. Li, Y. Zou, T. Cai, H. Chen, *Laser & Photonics Rev.* **2023**, *17*, 2300202.

[18] M. Veli, D. Mengu, N. T. Yardimci, Y. Luo, J. Li, Y. Rivenson, M. Jarrahi, A. Ozcan, *Nat Commun.* **2021**, *12*, 37.

[19] D. Mengu, A. Tabassum, M. Jarrahi, A. Ozcan, *Light Sci. Appl.* **2023**, *12*, 86.

[20] J. Qi, Y. Wang, C. Pang, H. Chu, Y. Cheng, *ACS Photonics* **2022**, *9*, 11.

[21] L. Li, J. Zhang, Y. Hu, J. Lai, S. Wang, P. Yang, X. Li, H. Duan, *Laser & Photonics Rev.* **2021**, *15*, 2100198.

[22] Y. Wang, Y. Wang, A. Yu, M. Hu, Q. Wang, C. Pang, H. Xiong, Y. Cheng, J. Qi, *Adv. Mater.* **2025**, *37*, 2408978.

[23] Y. Wang, A. Yu, Y. Cheng, J. Qi, *Laser & Photonics Rev.* **2024**, *18*, 2300903.

[24] X. Luo, Y. Hu, X. Ou, X. Li, J. Lai, N. Liu, X. Cheng, A. Pan, H. Duan, *Light Sci. Appl.* **2022**, *11*, 158.

[25] Y. Li, J. Li, Y. Zhao, T. Gan, J. Hu, M. Jarrahi, A. Ozcan, *Adv. Mater.* **2023**, e2303395.

[26] Y. Wang, C. Pang, J. Qi, *Laser & Photonics Rev.* **2024**, *18*, 2300832.

[27] R. Zhao, Q. Wei, Y. Li, X. Li, G. Geng, X. Li, J. Li, Y. Wang, L. Huang, *Laser & Photonics Rev.* **2023**, *17*, 2200982.

[28] X. Zhang, L. Huang, R. Zhao, H. Zhou, X. Li, G. Geng, J. Li, X, Li, Y. Wang, S. Zhang, *Sci. Adv.* **2022**, *8*, eabp8073.

[29] Y. Wang, C. Pang, Q. Wang, Y. Mu, Y. Cheng, J. Qi, *IEEE Trans. Microwave Theory Tech.* **2023**, *71*, 9.

[30] H. Lu, J. Zhao, B. Zheng, C. Qian, T. Cai, E. Li H. Chen, *Nat. Commun.* **2023**, *14*, 3301.

[31] Z. Liu, D. Wang, H. Gao, M. Li, H. Zhou, C. Zhang, *Adv. Photonics* **2023**, *5*, 034001.

[32] X. Hua, Y. Wang, S. Wang, X. Zou, Y. Zhou, L. Li, F. Yan, X. Cao, S. Xiao, D. Tsai, J. Han, Z. Wang, S. Zhu, *Nat. Commun.* **2022**, *13*, 2732.

[33] Y. Luo, D. Mengu, N. T. Yardimci, M. Veli, M. Jarrahi, A. Ozcan, *Light Sci. Appl.* **2019**, *8*, 112.

[34] X. Wang, S. Liu, L. Xu, Y. Cao, Y. Tao, Y. Chen, Z. Zhang, C. Chen, J. Li, Y. Hu, J. Chu, D. Wu, C. Wang, J. Ni, *Laser Photonics Rev.* **2023**, 2300880.

[35] E. Arbabi, A. Arbabi, S. M. Kamali, Y. Horie, A. Faraon, *Optica* **2016**, *3*, 6.

[36] J. Sisler, W. Chen, A. Zhu, F. Capasso, *APL Photonics* **2020**, *1*, 056107.

[37] M. Khorasaninejad, A. Zhu, C. Roques-Carmes, W. Chen, J. Oh, I. Mishra, R. Devlin, and F. Capasso, *Nano Lett.* **2016**, *16*, 11.

[38] S. Wang, P. Wu, V. Su, Y. Lai, M. Chen, H. Kuo, B. Chen, Y. Chen, T. Huang, J. Wang, R. Lin, C. Kuan, T. Li, Z. Wang, S. Zhu D. Tsai, *Nat. Nanotech.* **2018**, *13*, 3.

[39] Y. Wang, Q. Chen, W. Yang, Z. Ji, L. Jin, X. Ma, Q. Song, A. Boltasseva, J. Han, V. M. Shalaev S. Xiao, *Nat. Commun.* **2021**, *12*, 5560.

[40] J. Li, Y. Yuan, G. Yang, Q. Wu, W. Zhang, S. N. Burokur, K. Zhang, *Laser Photonics Rev.* **2023**, *17*, 3.

[41] Z. Li, R. Pestourie, J. Park, Y. Huang, S. G. Johnson, Federico Capasso, *Nat. Commun.* **2022**, *13*, 2409.

[42] S. Shrestha, A. C. Overvig, M. Lu, A. Stein, N. Yu, *Light Sci. Appl.* **2018**, *7*, 85.

[43] Y. Hu, Y. Jiang, Y. Zhang, X. Yang, X. Ou, L. Li, X. Kong, X. Liu, C.-W. Qiu, H. Duan, *Nat. Commun.* **2023**, *14*, 6649.

[44] Q. Jiang, J. Liu, J. Li, X. Jing, X. Li, L. Huang, Y. Wang, *Adv. Optical Mater.* **2023**, *11*, 15.

[45] S. Wang, P. C. Wu, VC. Su, YC. Lai, C. Chu, JW. Chen, SH. Lu, J. Chen, B. Xu, CH. Kuan, T. Li, S. Zhu, D. P. Tsai, *Nat. Commun.* **2017**, *8*, 187.

[46] W. Chen, A. Zhu, V. Sanjeev, M. Khorasaninejad, Z. Shi, E. Lee, F. Capasso, *Nat. Nanotech.* **2018**, *13*.

[47] W. Chen, A. Y. Zhu, J. Sisler, Z. Bharwani, F. Capasso, *Nat. Commun.* **2019**, *10*, 355.

[48] W. Ji, T. Cai, Z. Xi, P. Urbach, *Laser Photonics Rev.* **2022**, *16*, 2100333.

[49] Z.-B. Fan, H.-Y. Qiu, H.-L. Zhang, X.-N. Pang, L.-D. Zhou, L. Liu, H. Ren, Q.-H. Wang, J.-W. Dong, *Light Sci. Appl.* **2019**, *8*, 67.

[50] Y. Xu, J. Gu, Y. Gao, Q. Yang, W. Liu, Z. Yao, Q. Xu, J. Han, W. Zhang, *Adv. Funct. Mater.* **2023**, *33*, 35.

[51] H. Ren, J. Jang, C. Li, A. Aigner, M. Plidschun, J. Kim, J. Rho, M. A. Schmidt S. A. Maier, *Nat. Commun.* **2022**, *13*, 4183.

[52] E. Arbabi, A. Arbabi, S. M. Kamali, Y. Horie, A. Faraon, *Optica* **2017**, *4*, 6.

[53] X. Ou, T. Zeng, Y. Zhang, Y. Jiang, Z. Gong, F. Fan, H. Jia, H. Duan, Y. Hu, Tunable Polarization-Multiplexed Achromatic Dielectric Metalens. *Nano Lett.* **2022**, *22*, 24.

[54] A. McClung, M. Mansouree, A. Arbabi, *Light Sci. Appl.* **2020**, *9*, 93.

[55] Z. Li, P. Lin, Y.-W. Huang, J.-S. Park, W. T. Chen, Z. Shi, C.-W. Qiu, J.-X. Cheng, F. Capasso, *Sci. Adv.* **2021**, *7*, eabe4458.

[56] Q. Chen, Y. Gao, S. Pian, and Y. Ma, *Phys. Rev. Lett.* **2023**, *131*, 193801.

[57] W. Feng, J. Zhang, Q. Wu, A. Martins, Q. Sun, Z. Liu, Y. Long, E. R. Martins, J. Li, and H. Liang, *Nano Lett.* **2022**, *22*, 10.

[58] Y. Zhou, I. I. Kravchenko, H. Wang, J. R. Nolen, G. Gu, J. Valentine, *Nano Lett.* **2018**, *18*, 12.

[59] O. Avayu, E. Almeida, Y. Prior, T. Ellenbogen, *Nat. Commun.* **2018**, *8*, 14992.

[60] F. Balli, M. Sultan, Sarah K. Lami, J. T. Hastings, *Nat. Commun.* **2020**, *11*, 3892.

[61] C.-F. Pan, H. Wang, H. Wang, P. S, Q. Ruan, S. Wredh, Y. Ke, J. Chan, W. Zhang, C.-W. Qiu, J. Yang, *Sci. Adv.* **2023**, *9*, eadj9262.

[62]     H. Chung, O. D. Miller, *Opt. Express* **2020**, *28*, 5.

[63]     D. Sang, M. Xu, M. Pu, F. Zhang, Y. Guo, X. Li, X. Ma, Y. Fu, and X. Luo, *Laser Photonics Rev.* **2022**, *16*, 2200265.

[64]     M. Khorasaninejad, F. Capasso, *Science* **2017**, *358*, eaam8100.

[65]     M. Faraji-Dana, E. Arbabi, A. Arbabi, S. M. Kamali, H. Kwon, A. Faraon, *Nat. Commun.* **2018**, *9*, 4196.